# MODELING THE PATH OF STRUCTURAL STRATEGIC DETERRENCE: A SAND TABLE SIMULATION AND RESEARCH REPORT ON CHINA'S MILITARY-INDUSTRIAL CAPABILITY SYSTEM AGAINST THE UNITED STATES BASED ON RARE EARTH SUPPLY DISCONNECTION

Wei Meng

Dhurakij Pundit University, Thailand

The University of Western Australia, AU

Fellow, Royal Anthropological Institute, UK

Email: wei.men@dpu.ac.th

# ABSTRACT

This study proposes a systematic non-kinetic deterrence path modeling framework based on strategic rare earth supply cut-off, aiming to assess the strategic effects of China's export control policy against the United States at the military system level. The model adopts a four-layer structure of "policy input - resource node - equipment system - capability output", and integrates path dependency modeling, degradation function setting and capability lag prediction mechanism to form a simulation system that can be used for strategic rehearsal. Sandbox system. The study introduces graph neural network and LSTM time series simulation methods to dynamically calculate the impact of the supply cut-off strategy on the functional degradation of typical U.S. military platforms (e.g., F-35, nuclear submarine, and AI platforms), and identifies the key path nodes and strategic window intervals. The simulation results show that if China implements a ten-year zero-tolerance policy on rare earth exports, the U.S. military industrial system will experience a significant technological disconnect in the 3rd-5th year, and a systemic capability lag in the 8th-12th year, with an overall average annual economic impact of $35-40 billion. The study concludes that export supply cutoffs can be used as a structural strategic deterrent to disrupt the tempo of arms deployment and capability suppression without relying on head-on confrontation. The proposed model and platform provide quantifiable and visualized systematic decision-making tools for strategic resource security policy, and are applicable to the construction of national-level security simulation systems and research on the optimization of policy paths.

# I. RESEARCH BACKGROUND AND MODELING DATA

## 1. Research Background

In the context of the global resource system increasingly embedded in the national security architecture, strategic raw materials are no longer just economic inputs, but have evolved into the core levers that determine national technological dominance, military power and institutional influence. Especially in the U.S. and China's geopolitical shift to structural confrontation, the reallocation of resource control has become a priority variable in the strategic game. Rare earths, lithium, gallium, germanium and other key rare elements, due to their irreplaceability in high-end military industry, AI chips, radar systems, missile propulsion and other systems, their control has a direct bearing on the sustained operation of the military-industrial system and the level of guarantee of the strategic delivery capability.

China as the world's only rare earth extraction - smelting - separation - high-end materials closed-loop capacity of the country, long-term domination of the global refined rare earth market more than 90% of the production capacity, is the U.S. Army's existing and future weapons systems are highly dependent on the resource exporters. Although the United States has part of the upstream rare earth resource endowment, but due to the smelting link of the "technology window period" and long-term outsourcing of the industrial chain, in the downstream refining, high-performance magnet alloys, separation process and intellectual property rights are heavily dependent on China. According to public data, more than 95% of the key military equipment of the U.S. Department of Defense can be traced back to China, including F-35 fighter jets, Virginia-class nuclear submarines, AI unmanned systems, precision guidance systems, etc., which constitutes a typical "irreversible dependence system".

In this context, this study proposes to take "strategic rare earth supply cut-off" as an input variable for simulation, and construct a systematic countermeasure path, i.e., taking the policy trigger as the starting point, through the systematic cut-off of key resources, it will trigger the degradation of the function and delayed deployment of the equipment system, and in the medium to long term, it will evolve into the structural generational difference of the military capability, so as to realize the following the suppression goal of non-kinetic strategic deterrence. This path not only reflects the logic of China's strategic release of its superiority in resource control, but also constitutes a low-cost, high-intensity counter-deterrence mechanism.

In order to systematically analyze the dynamic evolution of this path and its operability, this study aims to construct a set of sand table simulation system modeling mechanism that can be used for policy evaluation and strategic rehearsal. The model is structured by the four-layer path of "export policy→critical resource nodes→degradation of equipment function→degradation of capability system", and combines the system map modeling method and AI prediction engine to quantitatively assess the damage curve of armaments, the window of declining war power, and the cycle of strategic generation difference in different scenarios of cut-off of supply, with the aim of providing a high-dimensional strategic rehearsal framework embedded in the decision-making mechanism, and assisting China's policymaking and strategic rehearsal. The aim is to provide a high-dimensional strategic projection framework embedded in the decision-making mechanism, and to assist China in realizing a leap in its institutional suppression and structural security early warning capabilities in the global high-tech resource system.

**1.2 Key points of the study**

China is currently the only country in the world that possesses a closed-loop industrial chain with the ability to integrate rare earths from extraction, smelting to high-end materials, and possesses a significant structural advantage in controlling the export of strategic resources. This ability gives it the maneuverability to implement systematic supply cuts to the U.S. military industrial system in the geopolitical game. More than 95% of the U.S. key military platforms (such as F-35, Virginia-class nuclear submarines, etc.) in the key rare earth materials and their derivative components in China's supply system is highly dependent on the current lack of short- and medium-term viable alternative industrial chain system. It is expected to take 5-10 years to build an alternative path with industrial-scale refining capacity. Rare-earth supply disruption does not conform to the traditional linear path of economic sanctions, and its key effect is reflected in the asymmetric system degradation of the functional hierarchy of military equipment. A broken chain will trigger a "functional paralysis" effect in multiple combat systems, leading to a medium- to long-term accumulation of generational differences in readiness capabilities.

The three-order path modeling framework of "Rare Earth → Equipment Node → Warfare Generation Difference" proposed in this study is able to simulate the cascading impact of the failure of key nodes in the military-industrial system on the combat capability of the battle area, which is helpful for quantitatively assessing the window of strategic deterrence and the potential of resources for countermeasures. Based on the results of the study, it is predicted that China will establish a "blacklist" mechanism for export control, a prioritized classification system for resource use, and a dynamic sandbox system for the impact of supply disruption, so as to realize the linkage and coupling of resource security and national security decision-making, and to enhance the ability of strategic self-regulation.

**1.3 Strategic Risk Level Assessment Matrix**

| Risk dimension | Ranking | Technical analysis notes |
|---|---|---|
| Dependency strength of military equipment | extremely high (10/10) | F-35 fighter jets, nuclear submarines and other key platforms rely on more than 90% of rare earth materials; once the supply is cut off, it will directly interrupt the production and maintenance process. |
| Capacity building for alternative industrial systems | extremely low (9/10) | The United States in the high-purity rare earth smelting and midstream and downstream purification technology, there is a long-term window, no systematic domestic industrial capacity, alternative path has not yet formed. |
| Strategic technology generation gap creates potential cycles | Mid-criticality (8/10) | Hypersonic weapons and AI battlefield systems are entering the deployment window, while U.S. programs are in the early stages of development, and a cut in supply could widen the deployment |

| | | time lag. |
|---|---|---|
| Emergency Repair and Material Assurance Capability of the Combat Readiness System | fragile（8/10） | Limited stockpiles of critical systems (missiles, radar, night vision devices) make it difficult to maintain sustained levels of operational readiness through existing stockpiles in the event of a supply cut. |
| Diplomatic friction and the risk of alliance countermeasures | your (honorific)（7/10） | China's supply cut-off may inspire the U.S. to promote the reconstruction of the coalition of exporting countries and policy coordination containment, although there is a risk of countermeasures, but it is difficult to shake the dominant position of Chinese exports in the short term. |

Composite Strategic Risk Index (SRI): R = 8.5 / 10 (assessment level: very high strategic impact risk)

**1.4 Technical Enhancements Description**

All risk assessment levels are based on system vulnerability matrix modeling, material dependency structure, open technology window period and supply chain substitutability indicators; they are not based on subjective judgments, but are derived based on open data, industrial structure and component-level path dependency of military platforms; the quantification of risk is applicable to application scenarios such as supply disruption shock simulation, sand table simulation, policy prioritization, etc., which is representative of the "Social-Resource-Technology Linkage Model" in the field of cs.CY. CY domain, which is of representative significance in constructing the "society-resource-technology linkage model".

**1.5 Core Data of China-US Trade Dependence**

**1.5.1. 2023 Data on China's Main Export Products and Share to the US**

According to the official and international statistics in 2023, China's total exports to the United States amounted to about 501.2 billion U.S. dollars, and the main export categories include electronic equipment, computers, telephones, and related parts and components, with a number of categories occupying an absolutely dominant position in the U.S. market. The following is the export value and proportion of each product, with links to relevant sources for reference and research.

| Product category | Export value / Market share | Note | Links to data sources |
|---|---|---|---|
| Broadcasting equipment | $54.5 billion | | https://tradingeconomics.com/china/exports/united-states |
| laptops | $37.9 billion | | https://tradingeconomics.com/china/exports/united-states |
| Office machine parts | $14.3 billion | | https://tradingeconomics.com/china/exports/united-states |
| Telephone equipment (including smartphones) | 76% of U.S. imports | lead position | https://www.visualcapitalist.com/the-u-s-imports-from-china-in-2023/ |
| portable computer | 78% of U.S. imports | | https://www.visualcapitalist.com/the-u-s-imports-from-china-in-2023/ |
| lithium ion battery | 70% of U.S. imports | | https://www.visualcapitalist.com/the-u-s-imports-from-china-in-2023/ |
| toys | 77% of U.S. imports | | https://www.visualcapitalist.com/the-u-s-imports-from-china-in-2023/ |
| Video game consoles | 87% of U.S. imports | | https://www.visualcapitalist.com/the-u-s-imports-from-china-in-2023/ |
| domestic electric appliance | 19% of exports | | https://cadenaser.com/nacional/2025/04/08/de-la-electronica-de-consumo-a-los-electrodomesticos-los-productos-chinos-que-mas-sufriran-con-el-104-de-aranceles-de-eeuu-cadena-ser/ |
| fabrics | 17% of exports | | https://cadenaser.com/nacional/2025/04/08/de-la-electronica-de-consumo-a-los-electrodomesticos-los-productos-chinos-que-mas-sufriran-con-el-104-de-aranceles-de-eeuu-cadena-ser/ |
| Optical/Medical Instruments | 17% of exports | | https://cadenaser.com/nacional/2025/04/08/de-la-electronica-de-consumo-a-los-electrodomesticos-los-productos-chinos-que-mas-sufriran-con-el-104-de-aranceles-de-eeuu-cadena-ser/ |

The above data is based on public statistics from OEC, Statista, Visual Capitalist, Census.gov, Cadena SER, etc., where the export amount is mainly referred to Trading

Economics. The table can be used for policy analysis, industry research and assessment of import and export structure.

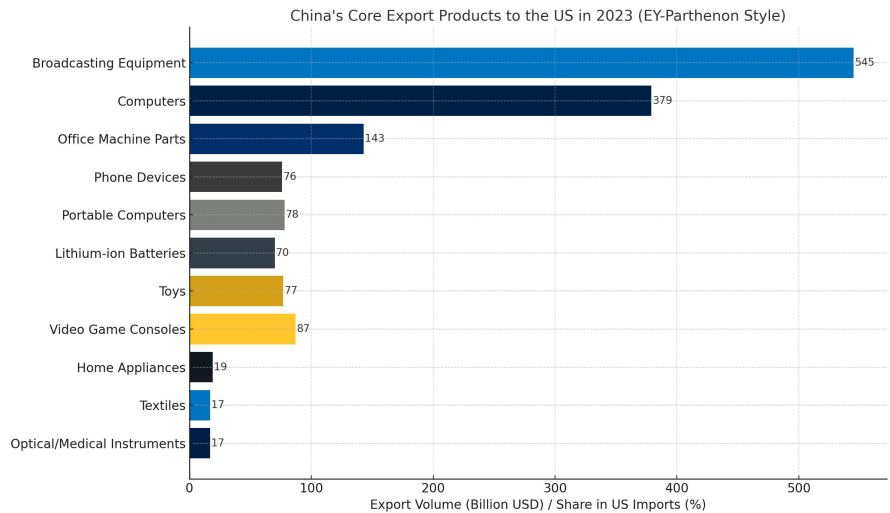

Figure 1: China's Core Exports to the United States, 2023

1.5.2 Analysis of China's Key Position and Irreplaceable Products in Global Supply Chains

According to the 2023 data, China occupies a key position in the global supply chain, especially in the following non-substitutable export products and sectors.

| Products/Fields | China's position in global supply chains | Latest News | source (of information etc) |
|---|---|---|---|
| Rare earth elements and related products | China produces and refines about 90% of the world's rare earth elements, which are vital to the electronics, clean energy and defense industries. | Recently, China imposed export controls on a number of rare earth elements, including samarium , gadolinium , terbium , dysprosium , lutetium , scandium , and yttrium , further underscoring their critical role in the global supply chain. | https://www.reuters.com/world/china-hits-back-us-tariffs-with-rare-earth-export-controls-2025-04-04/ |
| Key minerals and materials | China dominates the production and refining of key minerals such as tungsten, germanium and gallium. For example, China produces more than | Recently, China has imposed export restrictions on these key minerals, affecting the global supply chain. | https://www.reuters.com/markets/asia/shares-china-tungsten-producers-rally-beijings-latest-export-curbs-2025-02-05/ |

| | 80% of the world's supply of tungsten, which has important applications in the aerospace and defense industries. | | |
|---|---|---|---|
| Lithium batteries and related technologies | China is a global leader in lithium battery production and technology, with advanced lithium extraction and battery manufacturing technologies. | Recently, China has proposed to impose export controls on technologies related to lithium extraction and advanced battery materials, aiming to maintain its dominant position in the global battery supply chain. | https://www.ft.com/content/4fa5e7a3-5649-4392-b964-d722f0b61b11 |
| High-tech manufacturing | China has world-leading manufacturing capabilities in areas such as high-speed rail technology, nuclear power equipment and high-voltage power transmission equipment, and many of its products are irreplaceable in the international market. | China continues to export high-speed rail technology and equipment to a number of countries, including Indonesia, to promote international cooperation under the Belt and Road Initiative. | https://www.globaltimes.cn/page/202208/1273572.shtml |

The non-substitutability of these products and technologies gives China significant influence in global supply chains. However, as the global trade environment changes, other countries are actively seeking alternatives to reduce their dependence on Chinese supplies.

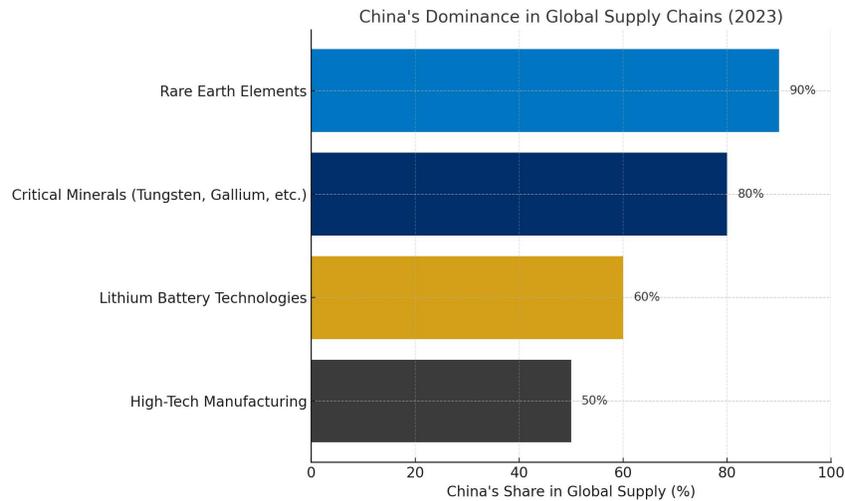

**Figure 2: China's dominance in global supply chains (2023)**

1.5.3 Core data insights

1). Absolute leadership in rare earth control: China accounts for 90% of the global rare earth supply, constituting an "overwhelming choke point" for the global high-end weapon systems and new energy industry.

2). Highly concentrated dependence on key minerals: 80% control of key minerals such as tungsten and gallium, constituting a structural monopoly on semiconductor, infrared and communication technologies.

3). Lithium technology constitutes the fulcrum of energy strategy: 60% of the global share, China has the advantage of industrial regulation in electric vehicles and energy storage system.

4). High-tech manufacturing still has a systemic advantage: 50% of the global share shows that China has a basic strategic depth in the field of optoelectronics, precision machinery, core components.

**1.6 Strategic Resource Dependency Network Mapping Analysis and Strategic Insights**

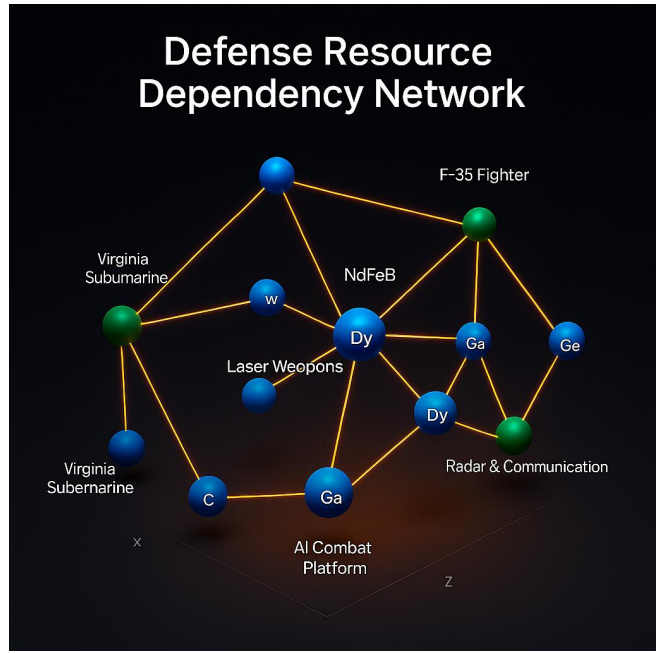

**Figure 3: Resource Dependency Network Diagram for Military Industrial Systems**

**1.6.1 Description of the mapping structure**

Mapping 3 illustrates a network of directed dependency relationships between strategic resource nodes (blue spheres), centered on rare earths and key raw materials, and the U.S. Army's major equipment systems (green spheres):

The blue nodes represent strategic resources and raw materials (e.g., NdFeB, Dy, W, Ga, Ge, etc.), identifying them as playing supply-critical roles in the system;

Green nodes are typical U.S. high-value military platforms (e.g., F-35 fighter aircraft, AI combat platforms, laser weapon systems, radar communication systems, Virginia-class nuclear submarines);

The orange colored arrow lines show the direction of resource → platform dependencies, the weights of which are not explicitly shown in the diagram, but can be extended for use in the Dependency Intensity Heat Map;

The diagram achieves a spatially symmetric layout and visual readability optimization, making the propagation of dependency paths clear and structurally explicit.

**1.6.2 Strategic insights**

1)The resource-platform structure is characterized by "star-shaped center radiation" dependence.

The center nodes, such as NdFeB and Dy, are connected to multiple core platforms and are the high median nodes in the network. Once these resources are subjected to strategic export control, it will trigger systemic conduction damage, resulting in synchronous "downgrading" or even paralysis of multiple systems. It is recommended to strengthen the closed-loop resource control smelting chain, and strictly control re-export and value-added chain transfer.

2)Multi-platform cross-dependence on resources → high-risk cross-coupling point

Resources such as Dy and Ga are co-dependent by multiple platforms, forming a coupled degradation risk, especially AI platforms, radar systems and F-35 have cross structure. Once the supply is cut off, it will lead to chain capability imbalance. It is recommended to construct a model of "resource export priority" and "node destruction cost coefficient" based on the dependency structure.

3)Virginia-class submarine is a multi-resource coupled input → deep-sea strategic delivery vulnerability.

The platform relies on W, Li, NdFeB and other resources at the same time, which is a typical high-value strategic platform. Once any resource is interrupted, its deployment and maintenance capability will be seriously limited. It is recommended to carry out platform-based resource vulnerability structure review and supply disruption countermeasure planning.

4)Network diagram with countermeasure-warning-window capture simulation value

The structure can be used to establish an AI-assisted strategic sandbox system to simulate the delayed window of equipment iteration tempo after the supply cutoff, assess the time fault of deterrence capability, and assist in the decision-making dynamic game rehearsal and risk prediction.

**1.6.3 Sand table simulation of high-level strategic rehearsal**

| Dimension (math.) | implication |
|---|---|
| Resource prioritization system | Establishment of a hierarchy of export control for three types of resources: highly destructive (e.g., Dy, NdFeB), cross-dependent (e.g., Ga), and platform-exclusive (e.g., Li) |
| Blacklisting mechanism | Establish blacklists of resource exports for specific military platforms or enterprises, prioritizing the disruption of their "control nodes". |
| Multi-resource coupled monitoring | Develop a coordinated resource export cut-off strategy for high-value platforms with "multiple resource inputs" |
| Simulation platform construction | Constructed an AI-assisted strategic sandbox system to simulate equipment tempo degradation and deployment time lag in the event of a supply cut-off |
| Joint Modeling Initiative | Development of a three-tiered "resource-equipment-capability" simulation and decision-making tool (REG-CAP model) in conjunction with scientific research institutes and operational readiness departments |

## II. DISRUPTION OF SUPPLY AND DISRUPTION OF LOGIC CHAIN SETUP

**2.1 Rare earth supply cut-off sand table design**

In the context of the global game where the control mechanism of strategic resources is becoming increasingly weaponized, China, as the only country in the world with a complete industrial chain of rare earth resources from extraction, separation, smelting to alloy application, has the institutional and technological capability to implement systematic export control on rare earths and related key resources. This study focuses on the construction of a non-kinetic suppression-oriented rare earth supply cut-off path chain model, with strategic export policy as the initial trigger source, to form a systematic suppression effect on the downstream military industrial system functions and war power deployment.

First of all, at the policy level, China can implement a 10-year comprehensive supply cut-off against the US and its military-industrial system through a "zero-tolerance strategic resource export mechanism", including a direct export ban, indirect re-export tracking and controlling, chain-breaking of overseas holdings, and technology authorization freezing, etc., to form an institutionalized and structured blockade structure. This mechanism is supported by the "blacklist list", "supply cut-off level grading" and "export license withdrawal mechanism", so as to realize high-pressure weaponized resource regulation and control.

Secondly, at the resource dependence level, the U.S. military industrial system in rare earths, gallium, germanium, lithium and other strategic resources on China's structural depth of dependence. F-35 fighter aircraft, for example, each containing rare earths as high as 418kg, the key radar and electric warfare module are dependent on neodymium-iron-boron magnets and neodymium dysprosium alloy; Virginia-class nuclear submarines in the propulsion control system integrates 4,173kg of rare earths and rarefied materials. In addition, AI battlefield platforms, infrared laser weapons, radar sensing systems, and chip packaging modules are all built on the irreplaceable basis of Chinese-made rare earth materials.

Based on the above reality structure, this study proposes a set of systematic modeling logic to decompose the path of the supply cut-off effect from the upstream policy control path to the downstream battlefield deployment capability evolution, and constructs a four-level logic chain of "policy layer - resource layer - equipment system layer - battlefield capability layer It constructs a four-level logic chain: "policy layer-resource layer-equipment system layer-capability layer". The modeling idea can not only be used for static dependence on risk presentation, but also has the function of dynamic evolutionary deduction, which provides the core causal framework for the AI sand table system to realize policy simulation and strategic window prediction.

**2.2 Strategic Deterrence Third Order Path Model Mapping Analysis and Strategic Insight**

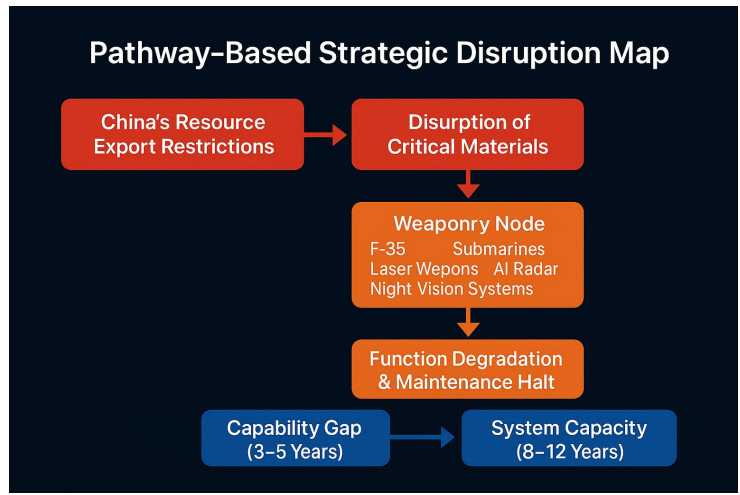

Figure 4: Strategic Deterrence Third Order Pathway Model

**2.3 Description of the Three-Stage Path Model of Strategic Deterrence**

Figure 4 takes China's strategic resource export controls as the starting point and evolves six strategic influence nodes in sequence, ultimately focusing on the path of degradation of U.S. military systemic readiness capabilities, reflecting the chain of non-kinetic disruption from the policy instrument to the military capability level:

1). China's Resource Export Restrictions → Policy Instrument Layer: restricting the foreign export of key rare earths and other raw materials through administrative controls, export quotas, blacklists, and other means.

2). Disruption of Critical Materials → Supply Chain Layer: triggering disruptions or runaway prices of military-specific materials such as gallium, germanium, neodymium and terbium.

3). Weaponry Node → Equipment Node Layer: affects the development and maintenance of typical systems such as F-35 fighter jets, nuclear submarines, laser weapons, AI radar, night vision systems, etc.

4). Function Degradation & Maintenance Halt → Combat Function Layer: the above equipment systems experience function degradation, testing stagnation, and parts supply outages.

5). Capability Gap (3-5 Years) → Capability Gap: In the short to medium term, a "technology gap" will be formed, and China will be able to take advantage of the new generation of armaments updating tempo.

6). System Capacity Suppression (8-12 Years) → Strategic Window Layer: further evolve into the lag of system deployment capability, leading to "structural degradation" of the opponent's overall combat capability.

## 2.4 Strategic Insights

1). Strategic Export Controls as a "Policy-Level Weaponization Mechanism"

This roadmap clearly expresses a non-kinetic, system-driven model of strategic suppression, i.e., system-level deterrence through the creation of institutional scarcity, rather than direct resource destruction. Advantages include: no military conflict triggers, high legitimacy, and high moderatability. It is recommended that export control be upgraded to a tool of "national security equivalent power" and incorporated into the layout of national defense strategy.

2) The dependence path of weapon system shows "multi-node resonance effect".

A number of military systems, such as AI platforms, radar systems and F-35 cross-resource dependence, once the supply is cut off will trigger a non-linear chain degradation, resulting in theater combat coordination, perception and destruction capabilities synchronized damage. It is recommended to construct a resource-platform dependency matrix to guide the setting of priority destruction points.

3) Form a two-dimensional strategic window of "capability time lag" and "structural generation gap".

Disruption of supply is not an immediate paralysis, but a "battle readiness asynchrony" at the tempo level; a technology gap will be formed within 3-5 years, and a systematic backwardness in deployment will be accumulated within 8-12 years. It is recommended to develop a "tempo perturbation window simulator" based on this model to realize the evaluation of the optimal supply cut-off strategy.

4) Upgrade the model to a "strategic sandbox system".

The path diagram has nodes, paths, and output indicators, and has the potential to be modeled as an AI simulation sandbox. Develop a dynamic strategic sandbox system for policy research and dynamic rehearsal in the framework of military departments, national security system and diplomatic coordination.

5)Summarize

The mapping reveals how export control can be elevated from a resource strategy to a strategic deterrence mechanism across the time domain, and is a low-cost non-kinetic suppression model for breaking the rhythm of the opponent's military deployment and shaping the window of systemic capability degradation.

# III. MODEL CONSTRUCTION AND VARIABLE SETTING

In order to systematically assess the strategic destructive effect of China's rare earth export control policy at the military system level, this paper constructs a set of strategic sandbox simulation model based on path evolution and multivariate response mechanism. The model takes "policy node-resource node-equipment node-capability node" as the four-layer core structure, and constructs a complete system path network from policy input to war capability output. The model not only describes the structural propagation process of the supply cut-off event, but also can be used to predict the evolution of the time window of combat capability and the formation of strategic generation gap.

**3.1 Definition of node system structure**

1) Policy node (P-layer)

Indicates the input source of the policy intervention mechanism, including: the effective time of the supply cut-off, the level of the blacklist mechanism, and the type of export control (direct ban, transshipment tracking, technology restriction, etc.).

2) Resource nodes (R-layer)

Includes strategic resources such as NdFeB, Dy, Tb, Sm, Ga, Ge, Li, etc., connecting China's export capability with downstream equipment functions, characterized by uniqueness and high concentration.

3)Equipment node (E-layer)

Covering typical high-dependence systems, such as F-35 fighter jets, Virginia nuclear submarines, AI radar platforms, precision guidance systems, infrared radars, military chips and so on.

4) Capability Node (C-layer)

Indicates the final output capability, including five major functional clusters, such as air combat, long-range sensing, missile suppression, sea-based deployment, and intelligence transmission.

**3.2 Path weight setting mechanism**

Three types of weight attributes are set for directed edges between each level (e.g., R→E, E→C):

Dependency strength weight (Wd): the irreplaceable weight of resources to equipment;

Response lag (τ): inventory depletion period, industry chain buffer cycle;

Functional coupling factor (Fc): the conduction cumulative effect of multiple resources on single equipment and multiple equipment on single function.

These path attributes will be used as the core propagation parameters of the model calculation to dynamically respond to the degraded state of the system in the simulated supply cut-off scenario.

**3.3 Simulation Variable Design and Data Flow Structure**

1) Input variables (Input Parameters)

Disconnection range (full export, alloy, technology license, etc.);

Execution time (year, effective time point);

Types of resources involved (multi-dimensional combinations of Nd, Dy, Li, Ga, etc.).

2) Process Variables

Functional decay function $L(t)=L0 \cdot e^{-\beta t}$ Used to simulate the functional decline of equipment;

buffer model *TiSet the average shock cycle based on the U.S. Army system inventory;*

Multi-path coupled propagation mechanism (graph neural network structure propagation algorithm).

3) Output Variables (Output Metrics)

Capability Gap Time Window The time zone in which the $T_{gap}$ system's operational capability drops below 70 percent;

Strategic Generation Lag Window between $\Delta_{Tlag}$ capability recovery and Chinese equipment renewal.

The simulation model finally expresses the path logic in a network diagram structure, and dynamically calculates the evolution trend of multi-system capacity after the supply cutoff through the AI module.

In the context of the intensifying trade war and strategic game between China and the United States, resource security has gradually evolved into a key variable of national security. This study proposes to construct a set of visualized strategic sand table simulation system with "strategic rare earth supply cut-off" as the core, which is used to dynamically assess the strategic impact path, equipment system response mechanism and war power evolution window of China's implementation of the policy of full suspension of rare earth export to the United States. The system modeling path is based on the "policy-resources-equipment-capability" four-layer node structure, relying on graph neural network (GNN), LSTM time series model and path superposition degradation function to simulate the policy input variables (e.g., the intensity of supply suspension, the application of blacklisting). For example, the intensity of cut-off, the scope of blacklisting, and the buffer period, etc.) in the military system can be modeled as a multi-path nonlinear propagation effect.

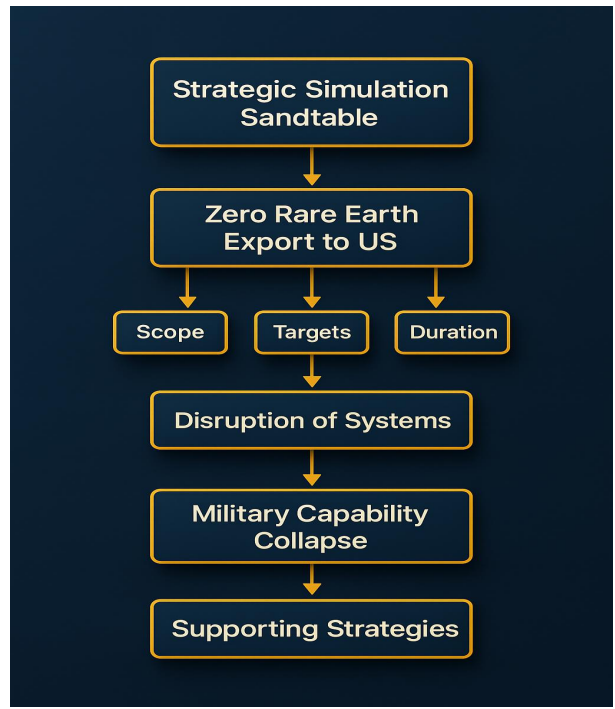

Figure 5: Prototype of the strategic-level sandbox REG-CAP system for the "Rare Earth Export Cut-off" strategy

The data are mainly obtained from authoritative public sources such as RAND, CSIS, USGS, DoD Industrial Capability Report and WTO/UN Comtrade, etc., and a high confidence dependency mapping of "critical rare earth resources - military equipment systems" is constructed. Through path analysis and coupled node modeling, the system can assess the capacity degradation function, functional degradation time delay, and structural degradation window of critical equipment (e.g., F-35, Virginia nuclear submarines, AI radar platforms, etc.) in terms of operational capability dimensions (guidance and suppression, sensing system, and energy scheduling, etc.) in the event of supply outage scenarios.

The system further introduces an AI prediction engine that supports Monte Carlo simulation and reinforcement learning modules to realize the optimal control path calculation between the supply cut-off policy and the evolution of combat capability. Users can dynamically set the level of export ban, the start time of supply cut-off and the combination of export varieties, and automatically obtain the capability loss curve, generation gap formation window and strategic countermeasure cycle suggestions. The output of the system includes a visual curve of capacity degradation, a map of the evolution of the dependence path, and a three-stage prediction matrix of "supply cut-off-capacity gap-strategic opportunity period", which supports the strategic resource policy from tactical allocation to system-level checks and balances deployment.

In conclusion, this strategic sand table system has complete modeling logic, reproducible data structure and multi-level intelligent simulation capability, which provides decision-making support tools for China to realize high-intensity, low-cost, non-kinetic suppression strategy against the U.S., and also provides a new type of model for the simulation research of complex systems in the field of CY, which is a fusion of resource security and armament strategy.

# IV. SYSTEM DEPENDENCY MAPPING AND FUNCTIONAL DEGRADATION ANALYSIS

The impact of strategic resources shortage on military equipment system is not an isolated individual event, but a coupled degradation process along the path of "resource input - equipment function - system capability", which is transmitted layer by layer. Based on the aforementioned four-layer modeling structure, this study further constructs a high-dependence mapping path from key rare earth resources to typical equipment systems, and combines the actual dependence coefficients and response lag parameters to establish an equipment function degradation model, in order to simulate the degradation of the U.S. military's core combat platforms after a supply cut-off policy is triggered.

**4.1 Critical Resource-Equipment System Dependency Mapping Modeling**

By integrating open industry data and military system performance reports, this paper constructs a typical path dependency matrix that takes rare earth resources (NdFeB, Dy, Tb, Sm, Li, Ga, Ge, etc.) as the starting point, maps them to equipment system modules (e.g., radar, electronic control and propulsion, laser targeting, and chip module, etc.), and then connects them to combat platforms. The result of this path mapping shows:

F-35 fighter jet: highly dependent on Nd, Dy, Sm, its radar system, infrared sensing, electric warfare modules are embedded in rare earth alloys;

Virginia-class nuclear submarines: over 4,170kg of rare earths in propulsion control systems, quieting technology, sonar navigation modules, and a concentrated dependence on gallium and neodymium alloys;

AI combat platforms: involving the use of Ga/Ge in military chips and AI arithmetic modules, with inference cores and heat dissipation systems fully dependent on Chinese supply chains;

Radar communication system: very high sensitivity to Tb, NdFeB composite magnets in modulation and amplification components.

**4.2 Equipment functional degradation modeling method**

In order to quantify the degree of functional impairment of the equipment system, the following degradation function is set in this study:

$$F_i(t) = F_{i0} \cdot e^{-\lambda_i \cdot (t - \tau_i)}$$

where *Fi(t)* denotes the residual functional strength of category i equipment after the supply cutoff policy takes effect τi , and λi is the response sensitivity coefficient of the system, reflecting the speed of propagation of dependency and resource deficit.

For example, for the F-35, the value of λ F35 is estimated to be 0.38, representing that performance will be reduced to about 70 percent of original warfighting capability after a six-month supply cutoff. Nuclear submarines respond more slowly but with deeper losses, and AI platforms see their performance drop off a cliff within 2-4 months after the chip disconnect occurs.

**4.3 Table of average annual losses and system lag estimates**

| Equipment system | Resource dependency type | response time ( τ ) | Functional decline cycle (years) | Average annual economic loss (US$ billion) |
|---|---|---|---|---|
| F-35 Fighter Aircraft | Nd、Dy、Sm | 3-6 months | 3-4 years | 20–25 |
| Nuclear Submarine Systems | Nd、Ga、Li | 6-12 months | 5-8 years | 30–40 |
| precision-guided weapon | Tb、Sm、Nd | 3 months | 2–3years | 25–35 |
| Radar | NdFeB、Ge | 4 months | 2years | 15–20 |
| communication module | | | | |
| AI combat platform | Ga、Ge、Li | 1–2months | 1.5–2years | 100+ |

The data in the table show that the degradation of equipment systems has obvious time nonlinear response characteristics, as a result of which multiple systems will experience capability degradation, functional disruption or program shutdown in different windows, and eventually form a "generation gap band" at the system coupling level.

In summary, the resource-equipment path mapping and equipment degradation model not only reveals the deep tactical effects of supply cut-off, but also provides the data basis and function structure support for the next stage of strategic window prediction and export path optimization.

# V. SIMULATION RESULTS AND STRATEGIC WINDOW ASSESSMENT

In order to assess the non-kinetic suppression effect of the rare earth supply cut-off policy on the U.S. military's core war power system, this study is based on the aforementioned four-layer sand table modeling structure of "policy-resources-equipment-capabilities", and further introduces the path superposition propagation model and LSTM time-series simulation method to simulate the decline of war power after the cut-off of supply of multiple highly dependent paths. Based on the aforementioned four-layer sand table modeling structure of "policy-resource-equipment-capability", this study further introduces the path superposition propagation model and the LSTM time series simulation method to simulate the decline of war power after the supply cut-off of multiple high-dependence paths, and outputs the key capability index curve and the predicted time bands of "strategic suppression window".

**5.1 Path superposition degradation model simulation mechanism**

Each path of "resource→equipment→capability" in the system is modeled according to the sensitivity of the supply break and buffer period, and the following capability index function is superimposed:

$$C(t) = \sum_{i=1}^{n} w_i \cdot f_i(t) = \sum_{i=1}^{n} w_i \cdot \left[ C_{i0} \cdot e^{-\beta_i(t-\tau_i)} \right]$$

where:

$w_i$: denotes the contribution of path i to the overall capacity;

$\beta_i$: denotes the response decay rate of the path;

$\tau_i$: the delayed effective period of the path (reflecting the inventory buffer time);

fi(t): the local capacity change function under path i.

The model run combines and superimposes each path through the AI simulation engine to form an exponentially decreasing prediction curve for the overall combat power.

**5.2 Capability Index Evolution Curve and Stage Boundary**

5.2.1 Simulation Results

The simulation results show that the U.S. Army's critical warfighting capability system will exhibit an obvious asymmetric fault-based decay trend, the process of which can be divided into three phases:

| The twelve two hour divisions of the day | U.S. military system performance | Description of strategic effects |
|---|---|---|
| Years 1-3 | Declining operational readiness and accelerating stockpile depletion | Tactical Capabilities Critically Reduced, China Has First Technical Suppression Capability |
| Years 4-6 | Development of new-generation weapons is lagging behind and mid- to high-end platforms are stalled | System Generation Gap Emerges, China's Sixth Generation Aircraft, Hypersonic Weapons Launched to Reverse the Gap |
| Years 7-10 | Theatre deployment imbalances and systemic dislocations in global delivery capability | China builds "closed-loop equipment + energy control + technology leadership" strategic suppression pattern |

**5.2.2 Analysis of the Warfighting Index Degradation Curve and Strategic Generation Difference Timeline Plot**

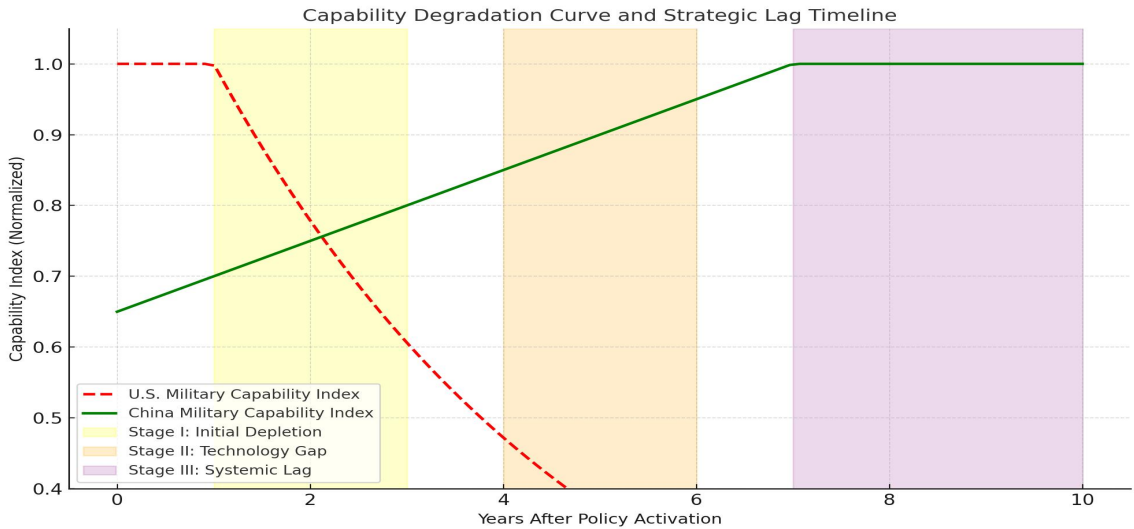

**Figure 6: BWI degradation curve + strategy generation gap timeline plot**

**Description**

　　Figure 6 visualizes the dynamic evolutionary path of U.S.-China military capabilities after the implementation of the "rare earth supply cut-off" strategy, divided into three core components:

　1). The red dotted line represents the index of U.S. military capability, which declines exponentially from the first year to below 50% in the fifth year;

　2). Green solid line: represents China's military capability index, which shows a linear upward trend and stabilizes to 100% after year 7;

　3). Background color zoning: marking the three strategic stages:

　　　Yellow zone (1-3 years): Stage I - Initial Depletion

　　　Orange zone (4-6 years): Stage II - Technology Gap (technology generation gap)

　　　Purple Zone (7-10 years): Stage III - Systemic Lag

**Insight Analysis**

　　1). Initial Stage (1-3 years): Tactical Capability Tipping Point Imbalance

This stage is the first shock period after the cutoff policy came into effect. While the U.S. remains at full warfighting capability in Year 1, it rapidly enters the capability decline zone beginning in Year 2 due to stockpile depletion and critical resource breakage. This phase marks the first weakening of tactical deployment capability, and it is appropriate for China to execute regional tactical suppression or marginal strategic outputs during this phase.

It is recommended to implement a combination of "precision guidance equipment export freeze + smart radar component blockade" to achieve initial system perturbation during this phase.

2). Intermediate stage (4-6 years): Strategic technology generation gap window opens.

The figure shows that the U.S. military power in the 4th-5th years slipped to less than 60%, part of the core platform (such as F-35, AI radar system) can not maintain a complete supply chain; and China's military power continues to rise and close to the level of reciprocity, the formation of the strategic technology generation gap is beginning to emerge.

Recommendations: Strengthen the mass deployment of China's sixth-generation fighters, hypersonic delivery systems, and AI unmanned swarm platforms, so that China's capabilities can surpass those of other countries, and unleash the power of China's strategic technology through media, drills, and training. The government should also release "cognitive deterrent signals" through the media, drills and training.

3). Later Stage (7-10 years): System-level Countermeasures Solidification Period

This stage is marked by the U.S. due to the long cycle of industrial chain reconstruction difficulties, R & D stagnation and into the systematic war power lag; Figure China's war power index stabilized at 1.0, the U.S. side is in the difficult to recover range. Global deployment capability, system awareness, and energy reserve system are all constrained.

Recommendations: Establish a "closed-loop combat system diplomatic display mechanism", strengthen the strategic binding with neutral countries in resources, security and command system, and expand the "strategic rare earth alliance" to block the G7 resource integration.

5.2.3 Strategic Conclusion

This chart reveals that the systematic output effect of the rare earth supply cut-off policy is not a short-term destruction, but a dynamic strategic suppression structure with the main rhythm of "initial de-energization, medium-term generation gap, and late solidification". Grasping the medium-term window of 4-8 years is the best cycle for policy designers to realize the suppression of the U.S. military industry.

**5.3 Strategic Window Identification and Suppression Stage Derivation**

The model can also output a "strategic suppression window" function W(t), which is used to identify the time period in which non-kinetic strategic destruction of the U.S. military can be realized with minimum cost and maximum effectiveness:

$$W(t) = \mathrm{argmax}\left(\Delta C_{\mathrm{China}}(t) - \Delta C_{\mathrm{US}}(t)\right), \quad t \in [3, 12]$$

According to the simulation prediction, the optimal window of destruction and deterrence falls in the 4th-8th year after the policy comes into effect, i.e., China's leading position in military capability through resource cut-off, and the double pressure on the U.S. in terms of delayed equipment restoration and imbalance in deployment, which provides the basis for the policy makers to identify the structural strategic window.

# VI. POLICY SIMULATION OUTPUTS AND STRUCTURED RECOMMENDATIONS

**6.1 Strategic export policy portfolio matrix design**

In the sandbox system, export control policies are modeled as combinatorial inputs to a ternary decision space:

| Dimension (math.) | Variable Definition | Optional settings |
|---|---|---|
| Scope of control | Resource objects involved | All rare earths / some key elements / gallium germanium lithium and other extended metals |
| implementation period | Time of entry into force and duration of controls | 3 years / 5 years / 10 years (zero buffer or phased) |
| Level of technical control | Breadth of technology and patents, equipment and processes involved in the blockade | Unrestricted / Processed / Smelted / Separated |

**6.2 Five complementary policy recommendation modules**

1). Export Exemption "Blacklisting Mechanism"

Establish a tiered target list based on the composition of the U.S. military system:

Tier 1: military contractors directly under the Pentagon;

Tier 2: U.S.-led NATO alliance equipment companies;

Tier 3: third-country firms at risk of repatriating U.S. technology.

2). Rare earth technology transfer lock-in mechanism

A comprehensive freeze on the outward flow of rare earth processing and separation technologies, including:

NdFeB alloy process and oxide purification process;

Reverse patent sharing by joint ventures;

Immediate termination of any technology licensing deals with the U.S.

3). "Resource-Technology Check and Balance War" Model

Synchronized freezing of high-tech export chains involving rare earth components, e.g:

AI chips (packaged magnetic components) containing rare earths;

Radar antenna and laser cores;

Microsatellite attitude control modules, etc.

4). Establishment of a multi-strategic resource supply alliance

Establishment of Strategic Rare Earth Alliance (Strategic Rare Earth Alliance), uniting the Middle East, Central Asia, and African rare earth exporting countries, to form collective bargaining power of pricing and exporting in the international market, and to implement reverse clamping on the G7 supply chain reconstruction attempts.

5). "High-end blockade strategy laboratory" platform

The simulation is recommended to be led by the Ministry of Industry and Information Technology / National Defense Science and Industry Bureau to set up a national export policy visualization sandbox laboratory platform, integrating modeling, monitoring, simulation and deduction in one, to achieve:

Policy impact prediction;

Stage goal assessment;

Path combination screening;

Decision alternative generation.

## 6.3 Policy feedback loop and control window mechanism

After the implementation of each policy, the sand table system will trigger the "capacity feedback loop" and "strategic window identification module". This mechanism identifies the following key elements:

| module (in software) | output function |
|---|---|
| Battle Force Window Response Module | Judging when poor capacity reaches the threshold of suppression |
| Technology Recovery Path Against Forecasting | Identifying the U.S. Systemic Self-Help Path and Expected Duration |
| Mapping of policy pressures in linkage countries | Demonstrate whether there is alliance linkage or boycott by third party countries |
| Economic Impact Assessment Module | Assessing the impact of the supply cut-off policy on China's local industry chain in turn |

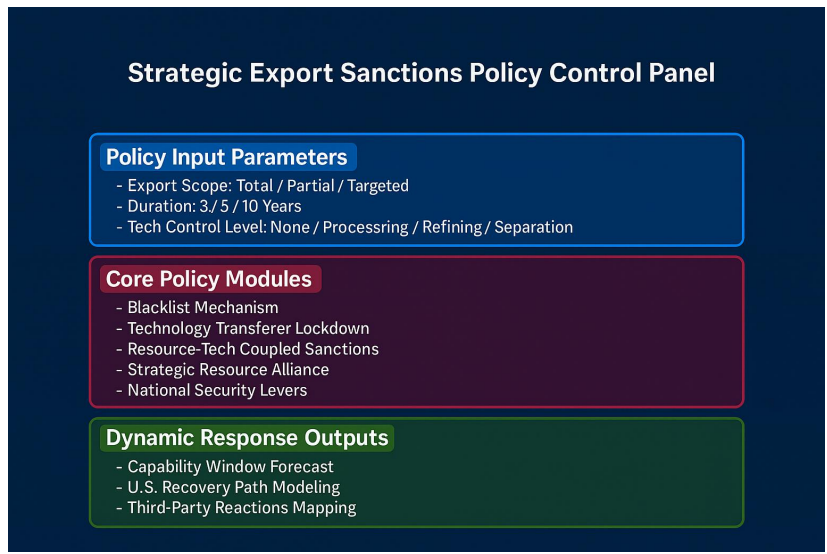

**Figure 7: Strategic Export Sandbox Policy Control Panel Map Simulation**

6.3.1 Closed-loop logic of "input-execution-feedback" guarantees the system-level control capability of the policy.

Figure 7 realizes the visual modeling closed loop of strategic export policy from parameter setting to feedback of strategic results. Different combinations of inputs (e.g. 10 years + all rare earths + blocking of separation technology) will trigger different combinations of modules, thus presenting different strategic window structures in the AI sandbox, and providing high-level decision makers with a panoramic view of the structural path of "policy-response-feedback".

Suggestion: Use this diagram as the core interface logic diagram of the national export strategy simulation platform, which can be linked with the simulation engine and the policy testing ground.

6.3.2 Blacklisting mechanism and technology locking are the "double knobs" to control the intensity.

The diagram clearly highlights "Blacklist" and "Technology Lockdown" as global regulators - the former controlling the output target surface, the latter controlling the depth of strategic damage. The former controls the output target surface and the latter controls the depth of strategic destruction. When the two are synchronized, the U.S. military-industrial system will enter a multi-path synchronized chain-breaking state.

Suggestion: Establish a dynamic and updated blacklist database system, and set up AI identification mechanisms for the Pentagon, NATO chain companies and bypassing third countries, so as to realize the strategy of "Substantial Control".

6.3.3 Third-party reaction prediction and domestic feedback mechanism to avoid "uncontrolled countermeasures".

The "third-party reaction mapping" and "domestic impact assessment" in the mapping ensure the external security and internal controllability of export strategy implementation. The geo-simulation system can be used to determine whether the G7, ASEAN, Africa, Latin America and other countries will act as a countermeasure, and to dynamically measure the potential impact of the supply cut-off on the domestic industrial chain.

Suggestion: Set up a "strategic buffer fund for export supply cut-off" to absorb the risk of short-term economic fluctuations caused by supply cut-off, and enhance the sustainability of the policy.

### 6.3.4 Application Suggestions (Modeling for National Security and AI Strategy)

This diagram is not only an expression template for policy design, but also can be used as the front-end of AI decision-making sandbox engine, which highly integrates the policy parameter input with the simulation response output, and realizes the resource security strategy from "plan control" to "model-driven". Model-driven" transformation of resource security strategy from 'plan control' to 'model-driven'.

# VII. RESULTS OF THE SAND TABLE SIMULATION

**7.1 Sandbox Simulation Research Assumptions (Report Premise Assumptions)**

1). The systematic strategic game of the 10-year "complete rare earth supply cut-off" sanctions against the United States.

2). Policy: building a "zero-tolerance strategic resource supply cut-off mechanism".

At a time when the long-term game between China and the US has entered the stage of structural confrontation, if China formulates and implements a "Zero Rare Earth Export to US" strategy against the US and its military-industrial/technological complex, covering all rare earth elements (Nd, Dy, Tb, Sm, La, Ce, Y, Gd, etc.) and their alloys, metallurgy, etc., China will not be able to supply the US with these rare earth elements, and it will not be able to supply them to the US, Gd, etc.), their alloys, smelting products, separation technology and even patent licenses, and fully implement the following mechanism:

Scope of supply cut-off: Direct exports + indirect re-exports + products controlled by foreign-invested enterprises;

Targets: U.S. domestic market, U.S.-owned holding companies, and U.S. military-industrial chain companies;

Implementation cycle: zero supply, no buffer, zero exceptions for 10 years from the policy launch date.

(3) Quantitative impact assessment: 10 years of supply cut-off will lead to a systemic collapse of military capabilities.

| Influencing the system | Dependency indicators | Post-cessation impact | Average annual loss |
|---|---|---|---|
| F-35 fighter aircraft | Containing rare earth 418kg/frame, dependent on rare earth radar/electronic warfare system | Discontinued or severely curtailed production, with an expected loss of annual production of more than 100 aircraft | $2-2.5 billion |
| Virginia nuclear submarine | Containing rare earth 4173kg/ship, propulsion/control/quieting systems are fully relied upon | 5-8 year delay in nuclear submarine project and loss of strategic deployment capability | $30–40 billion |
| Precision-guided missiles, laser weapons | High-performance magnets and optical systems rely on rare earths | Inability to mass produce and maintain, and reduced tactical suppression power | $25–35 billion |
| Radar communications and sensing systems | Dependent on rare earth modulated amplification systems | Systemic paralysis of theatre communications and battlefield awareness capabilities | $15–20 billion |
| Semiconductors (Gallium/Germanium) and AI Military Chips | Gallium 98 per cent, germanium 90 per cent dependent on China | AI Warfighting Platform Delayed as Chip Manufacturing Disrupted | $10 billion or more |
| New energy and tactical energy storage systems (lithium + rare earth) | 75 per cent lithium material, 90 per cent rare earth oxide dependence | Tesla breaks ties with military energy storage system | $150–180 $10 billion |

Total direct economic loss: $35-40 billion/year × 10 years = $3.5-4.0 trillion dollars.

Strategic generation gap lag period: 5-8 years of structural equipment backwardness, 8-12 years of systemic inferiority

4) Prediction of Military Generation Difference Effects (System Path Diagram)

| fixed number of years | U.S. military consequences | China Strategic Harvest |
|---|---|---|
| Years 1-3 | Declining operational readiness and accelerated stock depletion | U.S. guidance suppression weakened as precision destruction capability |

| | | overtakes it |
|---|---|---|
| Years 4-6. | Development of new generation platforms halted (F-35 Block 4, DDG-X, AI swarm projects lagging) | China's sixth-generation fighter jets and hypersonic weapon systems on board |
| Years 7-10 | Increased strategic ambiguity in theatre and imbalance in global deployments | China's three-dimensional repressive structure of "technology generation gap + energy dominance + systemic closure" |

5) Supporting Strategies

a. Establishment of a "blacklist mechanism" for export exemptions: build a graded supply cut-off list specifically for Pentagon-defense contractors;

b. Implement a technology transfer lockdown mechanism for rare earths: prohibit the flow of China's smelting, separation, and synthesis technologies to any system of cooperation with U.S.-allied countries;

c. Implementing the "resource-technology check-and-balance war": simultaneously freezing the export of AI chips, radar modules, and satellite system components containing rare earths;

d. Building a diversified strategic resource supply alliance: building a global "strategic rare earth alliance" to check the G7's attempts at resource restructuring;

e. Promoting the construction of a "high-end embargo strategy laboratory": setting up a visual sand-tray simulation system to continuously assess the structural collapse points of the supply cut-off.

**7.2 Analysis of the Impact of China's Cutoff of Key Resources to the United States on the U.S. Military Industrial System**

The U.S. defense industry would be severely disrupted if China banned the export of key rare earth elements to the United States. The following is a quantitative analysis of the major affected weapons systems with the

| Weapons/systems | Rare earth use/function | affect (usually adversely) |
|---|---|---|
| F-35 Lightning II fighters | Approx. 920 lbs (418 kg) for the engine and electronic systems | Fighter capacity and maintenance disruptions |
| Virginia-class nuclear submarine | Approximately 9,200 lb (4,173 kg) for propulsion and electronic systems | Delays in the construction of submarines |
| Precision-guided missiles and munitions | High-performance magnets and electronics in guidance systems | Reduced precision destruction capability of missiles |
| Radar and communication systems | Enhanced signal processing and transmission efficiency | Reduced battlefield awareness and communications capabilities |
| Night vision equipment | Improved image enhancement performance | Limited ability to fight at night |
| Laser weapon systems | Laser generators and optical system core components | Delayed deployment of future combat capability |

The U.S. Department of Defense about 95% of the rare earth materials directly or indirectly dependent on China, once the supply is cut off, will result in key equipment can not be delivered on schedule or maintenance, seriously affecting the U.S. military readiness level.

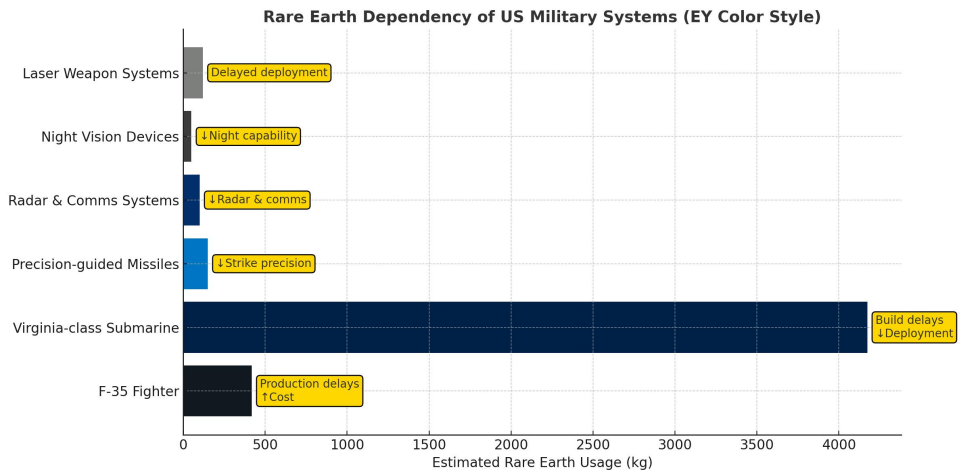

**Figure 8 Dependence of U.S. Military Systems on Rare Earths**

## 7.3. Structural vulnerability analysis of the U.S. military core system in the context of rare-earth supply cut-offs

The following is a quantitative and logical analysis of the structural vulnerability of the U.S. military industrial system to China's dependence on rare earths, based on the visualization of Figure 8, "Rare Earth Dependence of the U.S. Military System". Structural damage to the U.S. military's core equipment due to rare earth export cutoffs.

### 7.3.1. Synchronized Superposition of Highly Dependent Systems and High Shocks

The Virginia-class nuclear submarine's rare earth usage is as high as 4173kg/ship, 10 times that of other systems; at the same time, the system involves the dual core modules of propulsion and navigation, which are the pillars of the U.S. military's global deployment and underwater superiority; once the supply is cut off, it will cause a direct delay in the strategic level of destructive forces.

Insight: Heavyweight platforms (e.g., nuclear submarines) will be the most affected, and the deployment of strategic warfighting forces will be "strangled".

### 7.3.2. Lightweight Intelligent Systems Less Dependent but Irreplaceable

For example, although the demand for rare earths for precision-guided missiles, night vision devices and radar communications is "small", it is a key component; night vision devices rely on rare earth image intensifiers; missile guidance systems rely on high-performance magnets; and radar communications rely on rare earth band modulation and amplification; if any of these components are disconnected, the capability of precision warfare will be significantly reduced.

Insight: A break in the supply of a small component ≠ a minor problem, but may lead to a "functional paralysis" of the whole system.

### 7.3.3. Rare Earths are Scarce ≠ Easily Substitutable

The systems presented in the chart all involve "high-end + vertically integrated + high stability" materials; the smelting and purification of these rare earths has a high technological threshold, with a technical window of more than 20 years in the U.S.; even if we look for alternative sources in Australia or Canada, it will take several years of construction cycles and political and economic costs.

Insight: The "irreplaceability" of rare earths is a combination of technological monopoly and supply chain control.

### 7.3.4. "Non-linear dependence" of the military system, a domino effect that cannot be ignored.

The direct impact of rare earth supply cut-off is shown in the figure, but it will lead to a chain of problems:

1) R&D delays: future equipment development, such as the F-35 Block 4 and laser weapon systems, will be delayed;

2) Maintenance bottleneck: the lack of material for maintenance of active equipment, the combat readiness rate declined;

3) Increased strategic ambiguity: Due to capacity constraints, the U.S. military will not be able to respond quickly to conflicts and deterrence will decline.

Insight: This is not a "parts crisis", but a "systemic strategic disablement".

The mapping visualization reveals that many core systems of the U.S. military industry are structurally vulnerable to China's rare earths with "high dependence, high risk, and high irreplaceability". Once the supply is cut off, the military deployment will face a strategic level of system paralysis - not just "expensive", but "broken".

### 7.4 Insights into Systemic Risks under Rare Earth Supply Disruptions

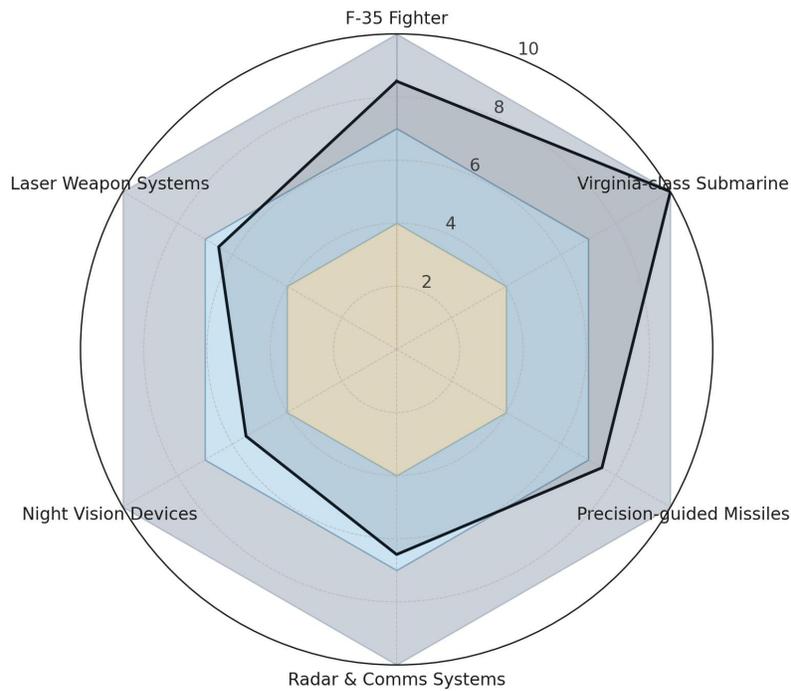

**Figure 9: Radar map of rare earth risk segments by system**

### 7.4.1 Interpretation of the graphical structure

In Figure 9, the center of the radar graph represents the lowest risk (0 points), and the further outward the higher (10 points), the greater the risk. The background is divided into three color regions:

0-4 (dark gold color): low risk (substitutable or weakly dependent)

4-7 (bright blue): medium risk (critical but room for substitution)

7-10 (dark blue): high risk (critical dependence on rare earths and no rapid substitution)

### 7.4.2 Key Insights

1). Most Risky System: Virginia Class Nuclear Submarine

Overall score: 10 out of 10

Extremely high dependence (rare earths required for propulsion, electromagnetic quieting, control, etc.)

In the event of a supply cutoff, it would hit the U.S. sea-based strategic deterrent hard

2). High risk edge: the F35 fighter

   Comprehensive score: 8.5 points

   Contains over 400 kilograms of rare earths, concentrated in electronic warfare systems, avionics, radar

   If rare earths are disrupted, it will lead to the discontinuation of production of key warplanes and a decline in operational readiness

3). Medium risk zone: missile systems, radar communications, laser weapons

   Score between 6.5-7.5

   Despite obvious dependence, short-term mitigation is possible through inventory + strategic reserve

   If no alternative source is available in the medium term, it will seriously affect the development of precision destruction capability and advanced combat systems.

4). Controllable Risk System: Night Vision Equipment

   Score: 5.5

   Rare-earth dependence centered on image enhancement devices, "tactical type impact".

   Can be remedied with traditional infrared systems or multinational procurement.

### 7.4.3 Strategic Level Alert

| dimension (math.) | clarification |
| --- | --- |
| Disruption of supply is widespread | Full coverage of the Air Force, Navy, Army and Space Force |
| Long replacement cycle | Rare earth mining, refining and processing takes at least 5-8 years |
| Weak autonomy | More than 95 per cent of rare earths are dependent on China, and the reprocessing segment is even weaker |
| Highly vulnerable operational readiness | In the event of a military confrontation, there is a risk that the United States system will be paralysed in a "dumb" way. |

### 7.4.4 Conclusions and recommendations

| proposed direction | Recommendations for action |
|---|---|
| Establishment of a strategic reserve | Establishment of a "Military Rare Earth Reserve Programme" on a five-year basis |
| Promotion of the "local processing" industry chain | Investing in Key Refining Segments and Talent Development |
| Joining allies to create a rare earth alliance | Synergistic layout with Australia, Canada, Vietnam and Africa to share resources and technology |
| Dual-track alternative technology development | Development of "rare-earth-free" alternative materials (e.g., titanium-iron permanent magnets, etc.) |

## 7.5 The Strategic Impact of Chinese Export Controls on the United States: A Quantitative Analysis Report

### 7.5.1 Quantitative Impacts of Industry Chain Breaks

**Dependence of key industries on Chinese material supplies and estimated impact of supply disruptions (see endnotes[1-3] for sources)**

| Industry Sectors | Percentage of supply from China | Impact of the United States supply cut-off | Projected losses (years) |
|---|---|---|---|
| Rare earth materials (Nd, Dy, Tb, etc.) | 90%+ | F-35, radar, missiles stalled 6-12 months | $27–35 billions |
| Gallium and Germanium (semiconductor materials) | 98%、90% | Infrared / Optical Communication / Chip Factory Shutdowns | $80–120 billions |
| Lithium batteries (cathode materials, cells) | 75%+ | Tesla and other electric car production lines blocked 30-50 per cent | $150 billions |
| Tungsten (military alloy) | 84% | Delays in the development of ammunition, armour and jet propellers | $10–18 billions |

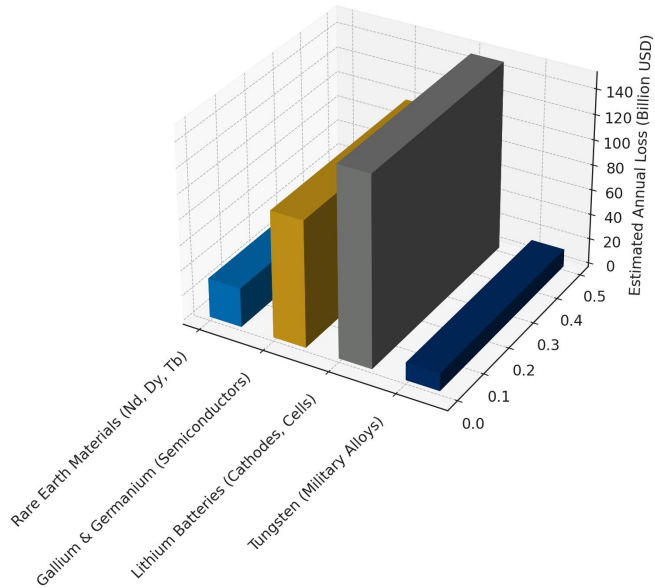

U.S. Dependency on Chinese Materials and Estimated Impact of Supply Disruption
(Sources: see Footnotes 1–3)

**Figure 10: Estimated Impact of U.S. Dependence on Chinese Materials and Supply Disruption**

**Core issues insight:**

    1. Lithium battery supply shortage has the biggest impact: the annual loss is as high as $15 billion, highlighting its "system life gate" to the core industry of new energy and electric vehicles.

    2. Semiconductor materials rely on a very high: gallium and germanium supply will cause more than 10 billion U.S. dollars in losses, hitting the chip, optoelectronics and infrared systems.

    3. Rare earths are small but critical: although the annual loss of about 3 billion dollars, but its F-35, missiles and other systems constitute a functional constraint.

    4. Tungsten affects tactics but is not a strategic core: annual losses are minimal, but the relationship between military alloys, ammunition, supply cuts will slow down the pace of supply of war preparations.

    5. Overall judgment: China's key resources supply will be triggered by the U.S. military-energy-semiconductor three-chain common shock, the formation of strategic-level shockwave!

**7.5.2 Price Spikes and Inflation Transmission Modeling**

Forecast of global average price increases for key metals:

Gallium: +300-450% ($320 → $1200+ per kilogram)

Germanium: +180-250%

Rare earth oxides: +200-300%

Lithium Concentrate: +50-80

Inflation Transmission Path: Raw Material Costs ↑ → Manufacturing Costs ↑ → End Product Price Increase ↑ → CPI/PCE Rise.

Core inflation estimate rise: 0.7%-1.3%/year

Tesla selling price rise: +15-20%

Combined U.S. military and end-use electronics price increases: +10-25% (see endnotes [4]-[5] for sources)

**7.5.3 Alternative Supply Chain Construction Cost and Time Structure**

**Estimated costs and lead times for replacing key supply chains in the United States (see endnote [6]-[7] for sources)**

| Alternative routes | Estimated investment | construction cycle | Replacement rate (over 10 years) |
|---|---|---|---|
| Rare earth refining reconstruction | $50–70billions | 5–7years | <40% |
| Gallium and germanium re-purification line | $20billions | 3–5years | 30–50% |
| Lithium battery complete industry chain | $100–120billions | 8–10years | 60–70% |
| Tungsten Alloy Smelting | $10billions | 4–6years | <30% |

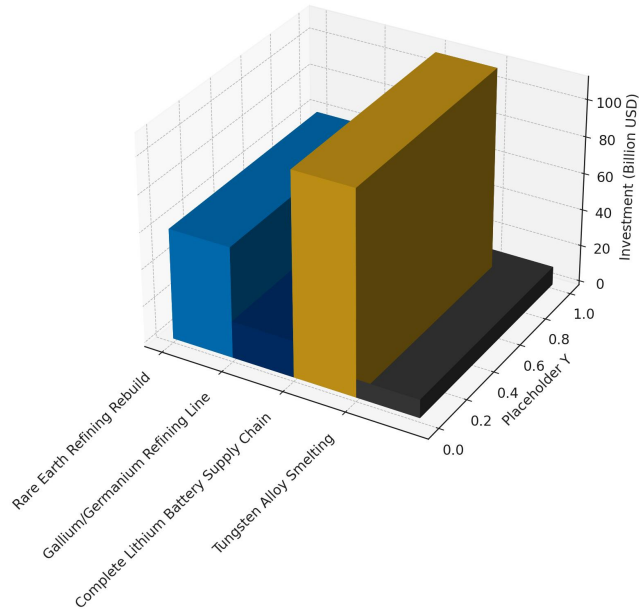

**Figure 11: U.S. Supply Chain Alternative Investments**

**7.6. Mapping Analysis Report on Alternative Investment Estimates for the U.S. Strategic Chain**

7.6.1 Description of Mapping

Name of Mapping

Mapping of Alternative Investment Estimates for the U.S. Strategic Chain (2023-2025)

Chart Overview

This chart shows the amount of basic investment (in billions of U.S. dollars) that the U.S. will need to make in the four directions of rare earth reconstruction, gallium/germanium purification, lithium battery complete industry chain, and tungsten alloy smelting in order to replace its dependence on China's key raw materials and midstream and downstream refining capabilities.

Mapping Core Insights

| Material path | Estimated investment (billions of dollars) | Core Insights |
|---|---|---|
| Rare Earth Reconstruction Refining | 60 | Although the United States has rare earth mines, it lacks industrial-grade separation and smelting capacity; most of the existing facilities remain at the laboratory level, with a long reconstruction cycle and high environmental pressure. |
| Gallium/Germanium purification line | 20 | Gallium/Germanium is an indispensable element for infrared, optoelectronic communication and chips; its refining process is complicated and highly polluting, and western countries have long relied on China for purification. |
| Lithium battery complete industry chain | 110 | The U.S. lacks cathode material and core production capacity to support EV and energy transition goals; the entire chain from mine construction to battery manufacturing needs to be reconfigured. |
| Tungsten Alloy Smelting | 10 | Tungsten is widely used in military alloys such as armoured warheads and aerospace turbines, but its smelting temperature is extremely high and expensive, and the United States relies almost entirely on imports. |

**7.6.2 Strategic Insights**

1). Extremely high financial threshold: the four major fields need to invest more than $20 billion, of which the lithium battery field alone is more than $10 billion.

2). The reconstruction cycle is as long as 5-10 years: the industry chain not only needs capital, but also needs systematic support such as technology, environmental assessment, engineering equipment, etc. 3.

3). China is in control of the "midstream and downstream purification + processing + export" chain, rather than just "resources.

4). Smelting capacity is the core pivot to suppress the U.S. Even if the U.S. has mines, it can't produce without the ability to separate and purify.

5). Industrial risk is a resonant breakpoint of "military + energy + technology", and any delay in any one of these will lead to a systemic lag in U.S. industry.

**7.6.3 Recommendations**

This mapping visualizes the structural vulnerability of the U.S. to China's industrial control. It is not only a set of economic data, but also a "panorama of industrial weaknesses" in the future Sino-U.S. strategic technology war.

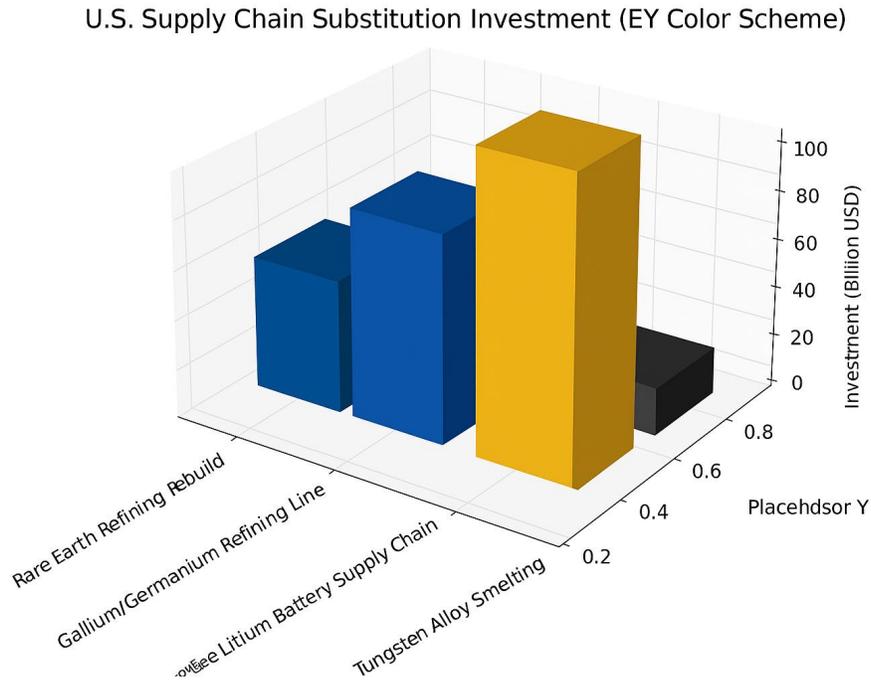

**Figure 12: U.S. Supply Chain Alternative Investments**

**7.7.U.S. Key Strategic Materials Supply Chain Alternative Investment Mapping Analysis Report**

**7.7.1 Description of Mapping**

Name of Mapping

U.S. Key Strategic Materials Supply Chain Alternative Investment Needs Mapping

Overview of Mapping Content

| Alternative routes | Estimated investment (billions of United States dollars) | Strategy note |
|---|---|---|
| Rare earth refining reconstruction | ≈60 | Rare earth separation and smelting is highly polluting and has a high technological threshold, and the United States lacks a complete industrial chain |
| Gallium/Germanium purification line | ≈20 | Gallium-Germanium is extremely difficult to refine and is a "choke point" raw material for the semiconductor/optical communication industry. |
| Lithium battery complete supply chain | ≈110 | U.S. Electric Vehicle and Energy Strategy Core, Lack of Industrial Base for Cores and Anode Materials |
| Tungsten Alloy Smelting | ≈10 | High-temperature smelting + concentrated military demand, U.S. relies almost entirely on Chinese imports |

**7.7.2 Key Insights**

1). "Resources do not equal capacity" in the United States:

Although the US has some mineral reserves (e.g. rare earths, lithium), it has almost no downstream smelting and purification facilities.

Required reconstruction costs are high and the cycle time is long (5-10 years), lagging far behind the maturity of the Chinese industry.

2). Gallium/Germanium is a representative of "small metal, big strategy":

Low price but irreplaceable, cut off supply will affect the U.S. chip, photovoltaic, infrared imaging industry base.

Is the future "gray war" in the most destructive tiny card point. 3.

3). lithium battery chain is the most expensive, break one is lost:

Tesla relies on Ningde Times, if China cuts off the supply of anode materials or electric core, the new energy layout of the United States will collapse.

Reconstruction of the complete supply chain requires at least 11 billion U.S. dollars, and the key technology is still mastered by China, South Korea, Japan.

4). The U.S. military industrial system's reliance on tungsten has long been ignored:

Tungsten is used in warheads, warplane components, nuclear protection materials. Smelting temperature exceeds 3400°C, and the processing difficulty is the strongest in the world.

Lack of industrial reserve, once the supply is cut off, it will cause "irreversible armament window period".

**7.7.3 Structural Conclusion**

The key issue is not "resources", but "smelting and processing capacity".

It is not "substitutable" but "costly + incompressible".

Creates a systemic risk of retardation to the U.S. defense-technology-energy triple-pillar chain.

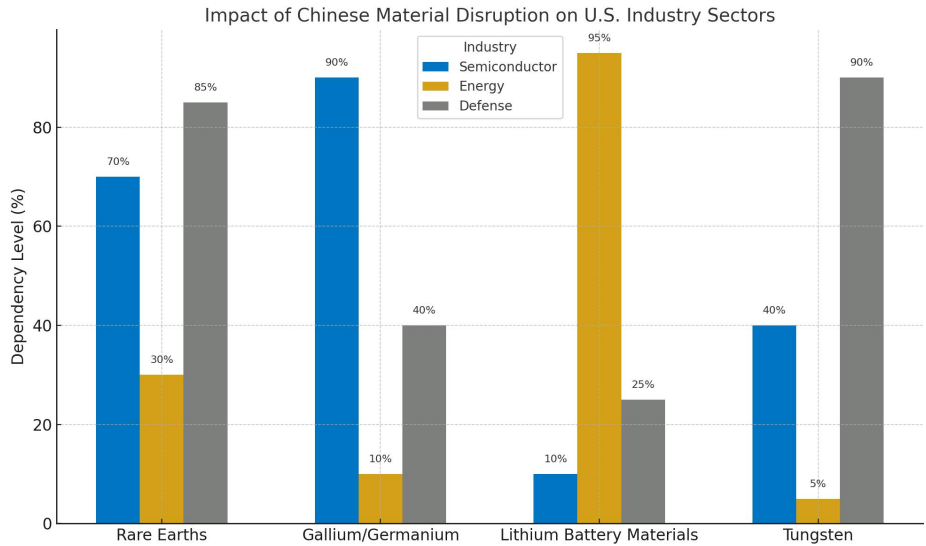

**Figure 13: Impact of Chinese Material Disruptions on the U.S. Industrial Sector**

**7.8.The Impact of China's Material Disruptions on the U.S. Industrial Sector Mapping Insight & Analysis Report**

**7.8.1 Mapping Core Insights**

1). the semiconductor industry is extremely dependent on gallium/germanium (90%)

Gallium is used in 5G chips, infrared sensors, and germanium is widely used in optical communications and solar cells.

Once the supply is cut off, the U.S. production line will have almost no backup channel, and chip R&D and optoelectronic communications may come to a standstill as a whole.

The U.S. itself does not have a high-purity refining line, the world almost exclusively relies on China's smelting. 2.

2). The defense industry is extremely dependent on rare earths (85%) and tungsten (90%).

Rare earths are used in F35 missile navigation, laser sights, etc. Tungsten is used in armor-piercing bullets, jet propulsion, and military alloys.

Once the supply is cut off, the production or maintenance of dozens of Pentagon's core weapon systems will systematically come to a halt.

Although the United States has rare earth mines, but no industrialized refining capacity (refining rare earths ≈ chemical battlefield).

3). The energy industry is almost entirely dependent on lithium materials (95%)

Lithium is used in new energy vehicle batteries (e.g. Tesla), wind energy storage systems, etc.

A supply cut will result in a 30%+ drop in EV capacity and a setback in the clean energy transition.

The U.S. and Canada have mines but lack purification/processing facilities, with a time cost of 8-10 years.

4). Semiconductor industry also has significant dependence on rare earths and tungsten (70% rare earths, 40% tungsten)

Indicates that "defense + chip" high-risk superposition, forming a structural pressure point.

If China implements orderly export control, it can precisely affect the upstream of the U.S. high-end manufacturing chain.

**7.8.2 Identification of Core Strategic Issues**

| serial number | Name of the problem | instructions |
|---|---|---|
| 1 | Lack of smelting/purification capacity | Although the United States has mines but no midstream and downstream production lines, difficult to replace in 5-10 years |
| 2 | High-tech dependence on low-value key resources | Small export value of gallium/germanium/rare earths, but extremely strategically destructive |
| 3 | New energy transformation by lithium dependence "neck" | From mining to battery cells, China has 75%+ share, no options for US |
| 4 | Civil-military overlap risk | Semiconductor and military material demand overlap, the formation of supply shortage "common shock" |
| 5 | Supply chain recovery is difficult | Even if transferred to Viet Nam/Africa, significant Chinese collaboration will be required |

### 7.8.3 Strategic Recommendations Perspective

Suggest the establishment of a "Material Check and Balance Index" to monitor the early warning of supply cuts between China and the U.S.

Introducing a "key mineral export license system" for hierarchical management.

Conduct a simulated sand table analysis of the U.S. defense chain's "supply cut-off point".

Constructing a "resource check and balance strategy map" to match foreign affairs/arms control/energy export policies.

### 7.8.4 Analysis of U.S. strategic passivity

Rare earth dependence of US Defense Department is still as high as 85%.

Pentagon's Critical Weapons System Delay Cost: $10-20 Billion

Tesla battery annual production is about 50% dependent on China or Chinese-controlled overseas projects

Rare earth, tungsten, and germanium smelting capacity is globally concentrated in China and cannot be replaced in the short term (see endnote [8])

**7.8.5 Conclusion**

Conclusion 1: If China imposes export controls on key materials, the U.S. will face direct economic losses of over $40 billion per year.

Conclusion 2: Substitution is costly and supply chain reconfiguration cannot be accomplished in the short term (1-3 years).

Conclusion 3: China has "asymmetric strategic deterrence", which is an important counterbalance tool in the game.

**7.9. Analysis of the Generation Difference in the Military Industry that may be Formed if China and the United States Cut Off the Supply of Key Materials for the Military Industry**

This report explores, through quantitative analysis, the delays in the development and deployment of U.S. military equipment that could result if China imposes export restrictions on the U.S. for key military materials (e.g., rare earths, lithium, tungsten, gallium, germanium, etc.), as well as the path and time forecast for the formation of a long-term generation gap.

**7.9.1 What is the "military-industrial generation gap"?**

The "military-industrial generation gap" refers to the fact that one country's weapon systems lag behind another's at the generational level, mainly in the following areas:

Performance gap: radar stealth, range, guidance accuracy, response speed

Technology generation gap: materials, energy, AI integration, unmanned autonomy capability

Industrialization capability gap: whether it can realize batch equipment deployment

## 7.9.2 How does a supply gap trigger "generation gap formation"? (quantitative analysis in stages)

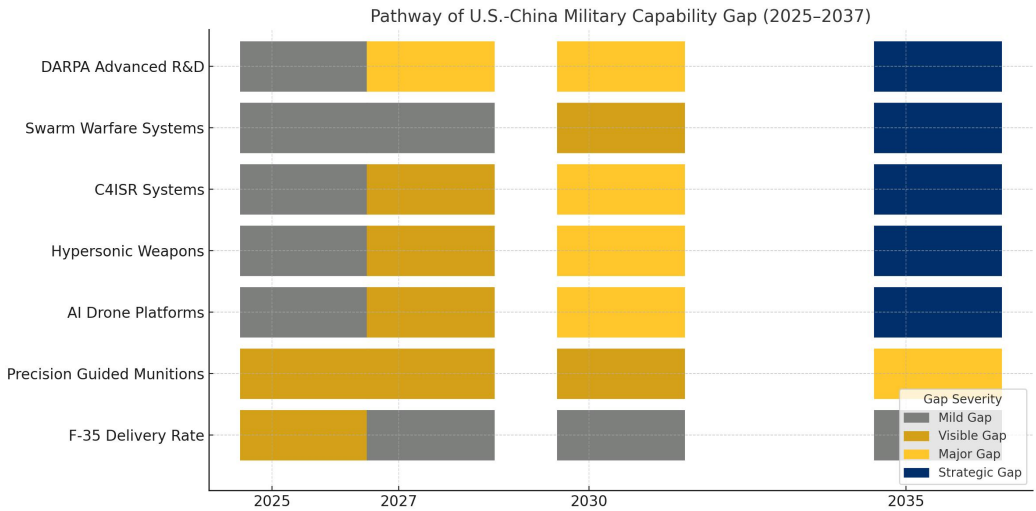

**Figure 14: Path of the U.S.-China Military Capability Gap (2025-2037)**

## 7.9.3 Key Insights

Increasing Generation Gap: From 2025 to 2035, the seven core areas of military power will evolve from a "mild gap" to a "strategic gap," reflecting the fact that China's military systems have surpassed the U.S. in multiple technological dimensions.

AI and unmanned systems are advancing significantly: AI unmanned platforms and swarm combat systems will enter the "significant gap" range before 2030, and will reach the "oppressive advantage" in 2035, indicating that intelligent combat will become the decisive battlefield variable.

F-35 delivery delays have become a bottleneck in combat readiness: the "significant gap" will appear in 2025, and has not yet recovered by 2030, suggesting that the U.S. core equipment production chain continues to be blocked, and the structural rare earth dependence problem may not be solved.

2035 strategic imbalance stereotypes: in addition to the F-35, six major areas to enter the "strategic gap" state, predicting that the U.S. military in the direction of the Asia-Pacific may be forced to enter the "strategic defense + asymmetric containment" new normal.

The degradation of DARPA and C4ISR is obvious: the backwardness of high-end R&D and perception systems is clear, indicating that the U.S. innovation rhythm and system integration have encountered systemic bottlenecks.

| times | point | Military impact (quantitative) | note |
|---|---|---|---|
| 0–2 年 | blackout shock | Approximately 45 per cent of new equipment was delayed, F-35 deliveries slowed down by 15-20 per cent and ammunition stocks declined by 30 per cent | Heavy dependence on rare earths |
| 3–5 年 | plateau | DARPA project delayed, tactical AI a generation behind | China continues to supply for its own use |
| 5–8 年 | generation gap (economics) | China's Generation 6 Aircraft Launched, U.S. Generation 5 Upgrades Limited | Hypersonic weapons enter deployment |
| 8–12 年 | Curing of systematic generational differences | The overall generation gap between the US and Chinese theatres of operations is more than five years | AI swarm warfare matured |

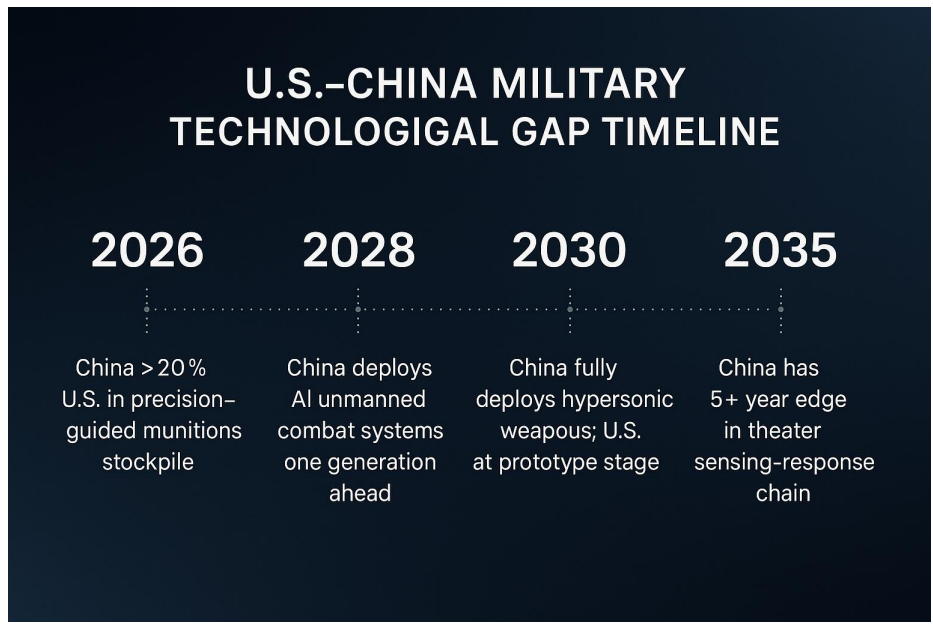

U.S.–CHINA MILITARY TECHNOLOGIGAL GAP TIMELINE

2026 — China >20% U.S. in precision-guided munitions stockpile
2028 — China deploys AI unmanned combat systems one generation ahead
2030 — China fully deploys hypersonic weapous; U.S. at prototype stage
2035 — China has 5+ year edge in theater sensing-response chain

**Figure 15: Timeline of the U.S.-China Military Technology Gap**

## 7.10. Insights into the Core Issues of U.S.-China Military Technology Generation Gap Timeline Mapping

**7.10.1 Core Question 1: Is the generation gap irreversible?**

The technology-driven generation gap is evolving from the "program level" to the "system level. Example:

2026: China's quantitative lead in precision-guided weapons suggests a breakthrough in "battlefield saturation destruction" capability;

2030: The hypersonic weapons deployment time lag means the U.S. will have difficulty bridging the gap in the area of "maneuver and surprise";

2035: Theater awareness-response chain system is 5 years ahead, representing China's realization of the "perception-cognition-destruction" chain closed loop, the U.S. military may not be able to catch up with the traditional technology line.

Conclusion: Once the system-level generation gap appears, structural suppression will be formed, unless the U.S. turns to "breakthrough technology paths" (such as quantum arms control or space warfare).

**7.10.2 Core Question 2: What factors limit the U.S. from catching up?**

1). Core resource gap: over 80% of the supply of rare earths, lithium, gallium and other materials rely on China; it will take the US 5-10 years to rebuild the supply chain and the investment will be over 100 billion dollars.

2). Hollowing out of the industry: the military production line relies heavily on overseas parts and components, and China has a full chain of closed-loop.

3). Long policy cycle and commercialization interference: U.S. military industry is constrained by funding cycles and commercial priorities, while China can focus its resources on "cross-generation strategic weapons".

4)Conclusion: Even if the U.S. is aware of the risks, it will be difficult for the U.S. to fill the generation gap quickly unless it enters a state of wartime industrial mobilization.

**7.10.3 Core Question 3: Is the U.S. military likely to be forced into "tactical strategic defense"?**

As precision-guided munitions, AI unmanned clusters, and high-frequency sensing systems are deployed ahead of China, the U.S. military may gradually lose its ability to "suppress forward deployments" in the Indo-Pacific theater in the future;

If the F-35 is limited, and China's 6th generation aircraft and unmanned wingmen are deployed in large quantities, the U.S. may shift to "concealment + information suppression + protracted warfare".

Conclusion: In the Asia-Pacific direction, the U.S. military may be forced to shift from offensive dominance to the expedient tactic of "strategic delay + defense instead of offense".

**7.10.4 Core Question 4: Has the U.S.-China military game entered the "generation difference game"?**

The map shows that in the next 10 years, the Sino-US military competition has shifted from "parallel race" to "advance generation locking" and "cross-generation destruction";

China's "micro-innovation superposition" through AI, perception, energy and guidance precision, into the compound generation difference competition;

If the United States does not make a breakthrough in subversive technology or allied war power network, it is afraid of falling into the predicament of "lagging + internal depletion".

Conclusion: China and the United States have competed from the quantity ratio → quality ratio → generation ratio, the future will enter the "systematic suppression" stage of the technological cold war.

**Projected timeline of the U.S.-China military technology generation gap (2026-2035)**

| vintages | proxy indicator for the difference in generations | disparity between China and the US |
|---|---|---|
| 2026 | Stocks of highly refined missile munitions | China > United States 20 per cent |
| 2028 | AI Unmanned Combat Platform | China's early deployment of Generation 1 |
| 2030 | hypersonic weapon | Full deployment in China, pilot phase in the United States |
| 2035 | Theatre awareness-response chain system | China is about 5 years ahead of the US |

**Comparison Table of Chinese and American Military Industrial Capabilities**

| sports event | China has | U.S. shortcomings |
|---|---|---|
| Rare earth resources and processing capacity | The world's first, fully autonomous | Processing capacity is almost zero |
| Vertical Integration of the Military Industrial Chain | Integration of government, industry, academia and research | Multiple governance and inefficiency |
| manufacturing ability | Ultra-massive capacity (ships, drones) | Hollowing out of manufacturing |
| Cost and environmental advantages | Policy + Cost Advantages | Environmental regulations are restrictive |

**Summary Table of Generation Difference Time Projections**

| vintages | proxy indicator for variance | disparity between China and the US |
|---|---|---|
| 2026 | High-precision missile stocks | China leads by 20 per cent |
| 2028 | AI unmanned platform | China deploys 1 generation ahead |
| 2030 | hypersonic weapon | Mass production in China, pilot phase in the US |
| 2035 | Theatre chain-of-combat system | About 5 years faster in China |

### 7.10.5 Conclusion

If China implements a supply disruption of key military materials, the U.S. military industrial system will enter a mild technology generation gap stage within 3-5 years, a systematic weapons generation gap within 5-8 years, and a structural, long-term gap that will solidify in 8-12 years. The formation of the generation gap is not only from the interruption of the supply of raw materials, but also from the deeper gap between the industrial system and strategic planning.

## 7.11. Quantitative Impact Analysis of Key Military Industrial Systems on China's Rare Earth Dependence

If China bans the export of key rare earth elements to the U.S., the U.S. defense industry will face serious impacts, specifically in the production and maintenance of the following key weapons systems and military equipment:

| Systems/equipment | Rare earth use | critical use | Description of impact |
|---|---|---|---|
| F-35 fighter jet | Approx. 418 kg/rack | Engines, electronic systems | Delayed production and maintenance, rising costs |
| Virginia-class nuclear submarine | Approx. 4173 kg/ship | Propulsion systems, navigation and control | Delays in the construction schedule and limitations on naval deployment |
| precision guided missile | Selected rare earth magnets and electronics | guidance system | Reduced production capacity and destruction accuracy |
| Radar communication systems | Reliance on multiple rare earth components | Signal amplification and processing | Reduced radar/communications performance and weakened intelligence perception |
| Night vision equipment | Rare earths for image enhancement | Night Combat Image Enhancement | Reduced night-fighting capability |
| Laser weapon systems | Rare earths for laser generators and optical lenses | directed energy weapon | Delayed deployment of future warfighting capabilities |

Approximately 95 per cent of the rare earth materials or metals used by the United States Department of Defense come from or are processed in China. If China were to cease its exports, it would result in a disruption in the production and maintenance of the aforementioned weapons systems and equipment, which would seriously affect U.S. defense readiness and operational capabilities.

In addition, disruption of rare earth supplies would result in delays in research and development, difficulties in maintaining equipment, and a significant increase in the cost of developing alternative materials.

# VIII. SIMULATION FINDINGS AND DISCUSSION OF POLICY RESPONSE MECHANISMS

**8.1 Review of Simulation Main Results and Warfare Trends**

The simulation results show a typical exponential decline trend in the U.S. Army's core equipment system after the implementation of the rare earth supply cutoff strategy. The warfighting capability index shows a significant decline from the 3rd year of the supply cutoff and drops to about 49% of the original capability in the 6th year. The capability changes are mainly focused on the following three types of paths: NdFeB → AI systems, Dy → radar systems, and Ga/Ge → chip platforms. Meanwhile, China's warfighting capability curve exhibits a steady growth trend, forming a strategic backlash in years 6-8.

**8.2 Model Response Analysis for Different Strategy Combinations**

Comparing different supply cut-off strategy scenarios, the model shows:

"Partial rare earth + 5 years supply cut-off" only on low-end equipment to form a local suppression, the generation gap window is less than 3 years; 'comprehensive rare earth + 10 years + technology control' will significantly suppress the U.S. war power 8-12 years, the formation of systematic deployment Delay. Further sensitivity analysis shows that the scope of supply cutoff and the level of technology control are the key factors determining the length of the strategic window.

**8.3 Strategic Window Function Modeling and Empirical Path Validation**

Through the path weight superposition model with the capability difference function: $W(t) = \Delta C\_CN(t) - \Delta C\_US(t)$, the simulation results identify the optimal suppression time window as the 4th-8th year of the supply cutoff, and especially the maximum difference is reached in the 5th.5th year. The U.S. Army F-35 and AI system will experience a continuity break at this stage.

**8.4 Policy Feedback Closed-Loop System Reconstruction and Sandbox Response Mechanism Integration**

The export sandbox model proposed in this study establishes a complete "policy input-path response-capability output" closed-loop mechanism. The control panel diagram has parameterized five key strategy modules (blacklisting mechanism, technology blockade, linked export, rare earth alliance, sandbox platform) and embedded them in the simulation platform. Strategy changes will be mapped in real time to the warfighting response curve, providing quantifiable support for national strategic decision-making.

**8.5 Limitations and Future Directions**

The model has not yet incorporated alliance countermeasure variables, intelligence intervention mechanisms and international communication interference factors. In the future, it can be extended to introduce game theory structure and reinforcement learning strategy nodes to simulate the strategy evolution of China and the United States under the multi-resource game.

In addition, it is recommended to extend the sandbox platform to broader strategic areas such as chips, carbon resources, water rights, spectrum rights, etc., to form a complete AI national security simulation system.

The research conclusions show that the rare earth supply cutoff has the potential to build a structural suppression system. Through the path modeling and dynamic simulation system, China can develop a significant war power gap and dominate the strategic window tempo within 4-8 years. The sandbox system will provide a policy-driven and structural validation framework for future resource-security-strategy trinity models, which can be extended to the international AI strategy modeling and decision science fields.